\documentclass{PoS}

\usepackage{german}
\usepackage{graphicx}
\usepackage{epstopdf}
\usepackage{amsmath,amssymb}

\newcommand{\beq}{\begin{equation}}
\newcommand{\eeq}{\end{equation}}
\newcommand{\bea}{\begin{eqnarray}}
\newcommand{\eea}{\end{eqnarray}}
\newcommand{\nn}{\nonumber}

\newcommand{\fig}{Fig.~}

\newcommand{\tr}{{\rm Tr}}
\newcommand{\bx}{{\bf x}}
\newcommand{\by}{{\bf y}}

\def\lsi{\raise0.3ex\hbox{$<$\kern-0.75em\raise-1.1ex\hbox{$\sim$}}}
\def\gsi{\raise0.3ex\hbox{$>$\kern-0.75em\raise-1.1ex\hbox{$\sim$}}}
\newcommand{\lsim}{\mathop{\lsi}}

\title{Baryonic or quarkyonic matter?}

\ShortTitle{Baryonic or quarkyonic matter?}

\author{\speaker{Owe Philipsen}, Jonas Scheunert\\ %\thanks{A footnote may follow.}\\
        Institut f\"ur Theoretische Physik, Goethe-Universit\"at Frankfurt am Main\\
        Max-von-Laue-Str. 1, 60438 Frankfurt am Main, Germany\\
        E-mail: \email{philipsen,scheunert@th.physik.uni-frankfurt.de} }
\abstract{
During the last years it has become possible to address the cold and dense regime of QCD directly for
sufficiently heavy quarks, where combined strong coupling and hopping expansions are convergent and a
3d effective theory can be derived, which allows to control the sign problem either in simulations or by fully
analytic calculations. In this
contribution we review the effective theory and study the $N_c$-dependence of the nuclear liquid gas transition, as well as the equation of 
state of baryonic matter in the strong coupling limit. 
We find the transition to become more strongly first order with growing $N_c$, suggesting that in the large
$N_c$ limit its critical endpoint moves to high temperatures to connect with the deconfinement transition. Furthermore,
to leading and next-to-leading order in the strong coupling and hopping expansions, respectively, the pressure is found
to scale as $p\sim N_c$. 
This suggests that baryonic and quarkyonic matter might be the same at nuclear densities. Further work is needed to see 
whether this result is stable under gauge corrections.
}

\FullConference{XIII Quark Confinement and the Hadron Spectrum - Confinement2018\\
		31 July - 6 August 2018\\
		Maynooth University, Ireland}

\begin{document}

\section{Introduction}

The QCD phase diagram plays a key role for various fields of current research in particle physics, heavy ion collisions and nuclear astrophysics.
Unfortunately, because of its complex action, i.e.~the ``sign problem'', lattice QCD with finite chemical potential for baryon 
number defies Monte Carlo simulations and most of the phase diagram remains unknown. Only the
low density sector with $\mu=\mu_B/3\lsim T$ is accessible with  
controlled approximate methods to circumvent this problem \cite{review}.  No sign of criticality is found there, and the transition from the
hadronic to the quark gluon plasma phase proceeds by an analytic crossover. This is in accord with the results of other non-perturbative 
methods applied in the continuum, such as Dyson-Schwinger equations \cite{cf}.
This situation motivates the development of 
effective lattice theories, whose sign problem can be overcome by algorithmic means or by altogether 
analytical treatment, such that also the interesting cold and dense region can be addressed. 
Here we briefly review such an approach and then discuss its extension 
to QCD with a general number of colours, $N_c$. This will allow us to establish contact with interesting conjectures about the
QCD phase diagram based on a large $N_c$ analysis, in particular the possible existence of a quarkyonic  phase \cite{quarky}.
For an early discussion of the $(T,\mu,N_c)$ phase diagram using various models and arguments, see \cite{mish}.

\section{An effective lattice theory for QCD with heavy quarks}

Starting point is lattice QCD with the standard Wilson action. Finite temperature gets implemented by a compact euclidean time dimension
with $N_\tau$ slices, $T=1/(aN_\tau)$, with (anti-)periodic boundary conditions for (fermions) bosons. 
An effective theory in terms of temporal links only
is obtained after integrating over the quark fields and gauge links in spatial directions in the partition function,
\beq
Z=\int DU_0DU_i\;\det Q \; e^{-S_g[U]}\equiv\int DU_0\;e^{-S_{eff}[U_0]}=\int DW \,\;e^{-S_{eff}[W]}\;.
\eeq 
With the spatial links gone, the effective action depends on the temporal links only via Wilson lines closing through 
the periodic boundary, or Polyakov loops,
\beq
W({\bf x})=\prod_{\tau=1}^{N_\tau} U_0({\bf x},\tau),\quad L({\bf x})=\tr W({\bf x})\;. 
\eeq
This effective action is unique and exact. The integration over spatial links causes long-range interactions 
of Polyakov loops at all distances and to all powers so that in practice truncations are necessary. 
For non-perturbative ways to define and determine truncated 
theories, see \cite{wozar,green1,green2,bergner}.
Here, we expand the path integral in a combined character and
hopping expansion, with interaction
terms ordered according to their leading powers in the coefficient of the fundamental character $u$
and the hopping parameter $\kappa$,
\beq
u(\beta)=\frac{\beta}{18}+\frac{\beta^2}{216}+\ldots < 1, \qquad \kappa=\frac{1}{2am_q+8}\;.
\eeq
The dependence of $u$ on the lattice gauge coupling $\beta=2N_c/g^2$ is known to arbitrary precision, and
$u$ is always smaller than one for finite $\beta$-values. 
Since the hopping expansion is in inverse quark mass, the effective theory to low orders is valid for heavy quarks only.
Both expansions result in convergent series within a finite radius of convergence.
Truncating these at some finite order, the integration over the
spatial gauge links can be performed analytically to provide a closed expression for the effective theory.

\begin{figure}[t]
\centering
\includegraphics[width=0.9\textwidth]{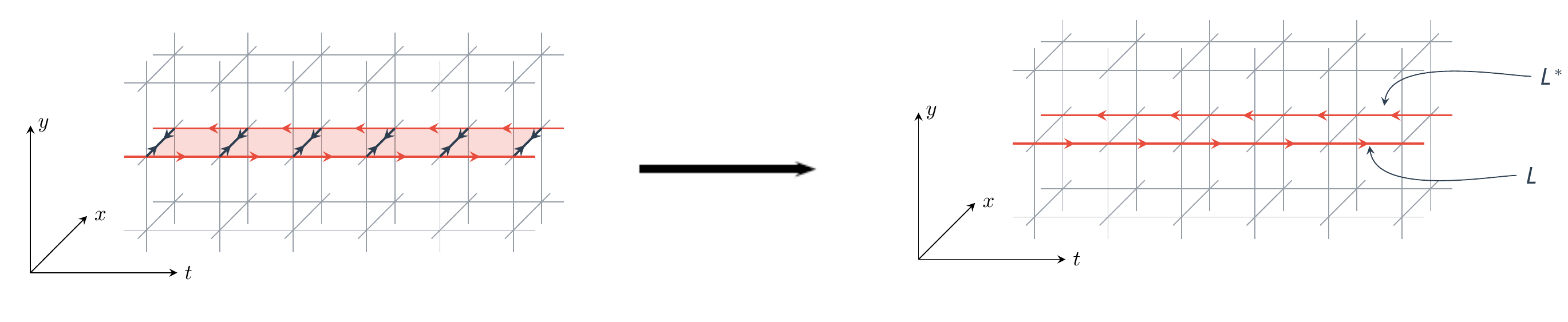}
\caption[]{Integrating over spatial links transforms the slice of lattice action into a nearest neighbour 
interaction of a Polyakov loop and the neighbouring conjugate Polyakov loop.}
\label{fig:zn}
\end{figure}
The procedure is illustrated for the leading order term of the pure gauge action in \fig\ref{fig:zn}.
Integrating over the spatial links of a chain of plaquettes along the temporal direction results in
\begin{eqnarray}
e^{S_{\mathrm{eff}}^{(1)}}=\lambda_1(u,N_\tau)\sum_{<\bx\by>}\left(L_\bx L_\by^\ast+L_\bx^\ast
L_\by\right)\;,
\qquad\lambda_1(u,N_\tau)=u^{N_\tau}\Big[1+\ldots\Big]\;.
%\lambda(u,N_{\tau}\geq5)&=&u^{N_\tau}\exp\bigg[N_{\tau}\bigg(4u^4+12u^5-14u^6-36u^7
%		+\frac{295}{2}u^8+\frac{1851}{10}u^9+\frac{1055797}{5120}u^{10}+\ldots\bigg)\bigg]\;,
\label{eq_lambda}
\end{eqnarray}
To leading order every plaquette contributes one factor of $u$ \cite{polon}. 
Higher orders consist in deforming the surface by cubic extensions
into the other dimensions and, together with a resummation leading to a logarithmic action 
and sub-leading couplings, can be found in \cite{efft1}. 
Going via an effective action
results in a resummation to all powers with better convergence properties compared to a direct series expansion of 
thermodynamic observables as in \cite{lange1,lange2}. 
Since the Polyakov loop $L({\bf x})$ contains the length $N_\tau$ of the 
temporal lattice extent implicitly, the effective theory is three-dimensional. Note that this representation of a 4d Yang-Mills theory
by a 3d centre-symmetric effective theory is the basis for the Svetitsky-Yaffe conjecture \cite{sy}.
 
Including the quark determinant via the hopping expansion introduces centre symmetry breaking terms and additional effective
couplings $h_i$ \cite{fromm},
\begin{eqnarray}
-S_{\mathrm{eff}}=\sum_{i=1}^\infty\lambda_i(u,\kappa,N_\tau)S_i^s-
2N_f\sum_{i=1}^\infty\left[h_i(u,\kappa,\mu,N_\tau)S_i^a+
\bar{h}_i(u,\kappa,\mu,N_\tau)S_i^{a,\dagger}\right]\;.
\label{eq_defseff}
\end{eqnarray}
The $\lambda_i$ are defined as the effective couplings of the 
$Z(3)$-symmetric terms $S_i^s$, whereas the $h_i$ multiply the asymmetric 
terms $S_i^a$. 
In particular, $h_1, \bar{h}_1$ are the coefficients of $L,L^*$, respectively, and
to leading order correspond to the fugacity of the quarks and anti-quarks,
\beq
h_1=(2\kappa e^{a\mu})^{N_\tau}(1+\ldots) =e^{\frac{\mu-m}{T}}(1+\ldots), \quad 
\bar{h}_1=(2\kappa e^{-a\mu})^{N_\tau}(1+\ldots) =e^{-\frac{\mu+m}{T}}(1+\ldots)\;
\eeq
with $m=-\ln(2\kappa)$ the leading order constituent quark mass in a baryon, while $h_2=\kappa^2 N_\tau/N_c(1+\ldots)$
is the leading order coefficient of an $L_\bx L_\by$ interaction term.
As an example, we give the partition function including just these simplest interactions,
\bea
\label{zpt}
Z&=&\int DW\prod_{<\bx, \by>}\left[1+\lambda(L_{\bx}L_{\by}^*+L_{\bx}^*L_{\by})\right]\\
&&\times\prod_{\bx}[1+h_1L_{\bx}+h_1^2L_{\bx}^*+h_1^3]^{2N_f}[1+\bar{h}_1L^*_{\bx}+\bar{h}_1^2L_{\bx}+\bar{h}_1^3]^{2N_f}
\nonumber\\
&&\times \prod_{<\bx, \by>}\left(1-2h_{2}\left({\rm Tr} \frac{h_1W_{\bx}}{1+h_1W_{\bx}} - {\rm Tr} \frac{\bar{h}_1W^\dag_{\bx}}{1+\bar{h}_1W_{\bx}}\right)\left({\rm Tr} \frac{h_1W_{\by}}{1+h_1W_{\by}} - {\rm Tr} \frac{\bar{h}_1W^\dag_\by}{1+\bar{h}_1W^\dag_{\by}}\right)\right)
\times \ldots \;.\nn
\eea
In this expression the first line comes from the pure gauge sector, the second line is the exact static determinant and the third line
the leading correction from spatial quark hops. This partition function has a weak sign problem and can be simulated with 
either reweighting or complex Langevin methods \cite{fromm,bind}.
Since the effective couplings are expressed by powers of expansion parameters they are themselves small, 
so the theory can be treated by linked-cluster expansion methods known from statistical physics \cite{k8},
with results in perfect agreement with the numerical ones. In this way, full control over the sign problem is achieved.

\subsection{The deconfinement transition at zero and finite density}

In full lattice QCD, for any choice of $N_\tau$, there are critical couplings 
$\beta_c(N_\tau), \kappa_c(N_\tau)$ marking the deconfinement transition at finite temperature. These critical values 
represent non-analyticities in the thermodynamic functions, and hence 
limit the radius of convergence for the respective series expansions. 
The effective theory thus gives a valid description of the confined phase only. Nevertheless, it can be used to find the location of phase 
transitions. In the effective theory, the deconfinement transition is represented as spontaneous centre symmetry
breaking at some configuration of critical couplings $\lambda_{i,c}=\lambda_i(u_c, \kappa_c, N_\tau), h_{i,c}(u_c,\kappa_c,N_\tau)$. 
The values of these can be determined by cheap numerical simulations of the effective theory. Inversion of the 
effective couplings as functions of the QCD couplings then gives predictions for 
$\beta_c(N_\tau), \kappa_c(N_\tau)$, which can be compared with the results from full QCD simulations. 

For $SU(3)$-Yang-Mills theory, the simplest effective theory with only a nearest neighbour coupling correctly 
reproduces the order of the deconfinement transition to be 3d Ising for $SU(2)$ and first-order for $SU(3)$.
Moreover, the predicted $\beta_c(N_\tau)$ are within 10\% of the true values for $N_\tau\in[2,16]$ \cite{efft1}.
For QCD with heavy quarks, the explicit centre symmetry breaking weakens the first-order phase transition until it
ends in a second-order critical point at some critical quark mass.  For the simplest effective theory including only the static determinant,
the predicted $\kappa_c$ agrees to better than 10\% with full QCD simulations on $N_\tau=4$ \cite{fromm}, 
finer lattices are not available at present. 

\begin{figure}[t]
\centering
\includegraphics[width=7cm,clip]{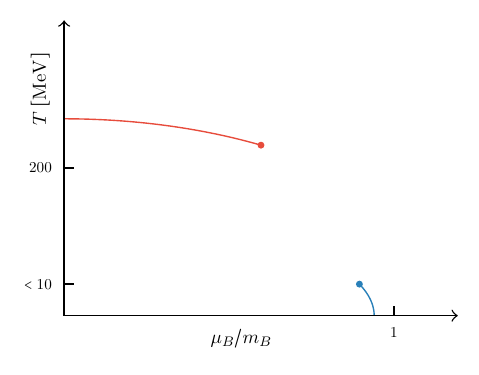}
\includegraphics[width=7cm,clip]{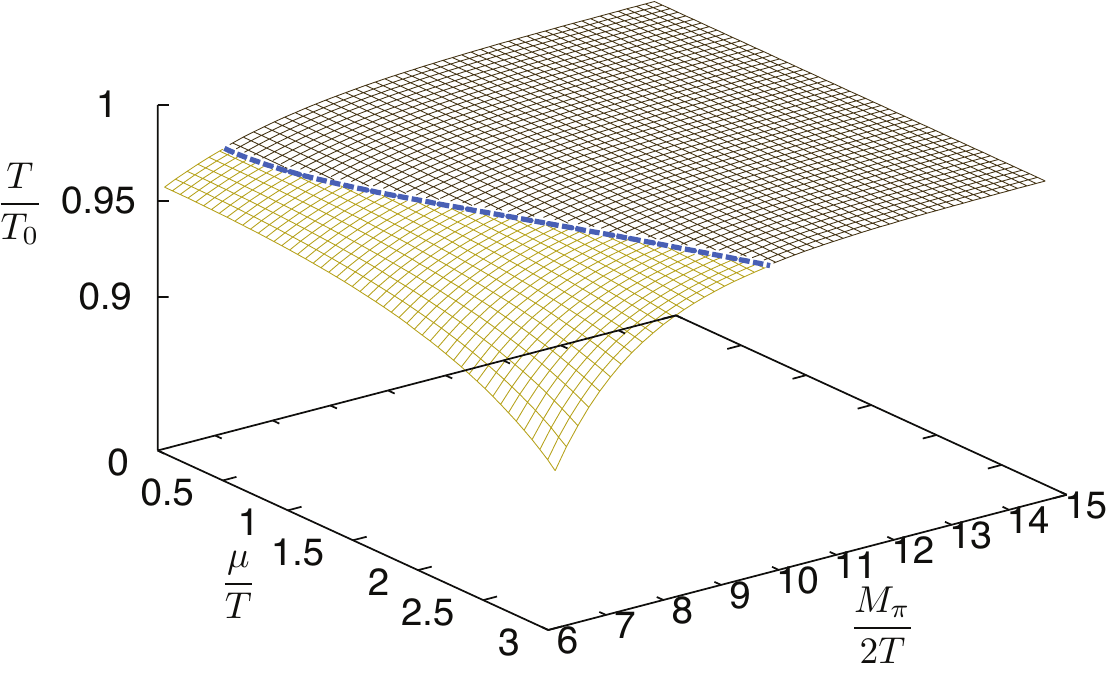}
\caption{Left: Qualitative phase diagram for QCD with very heavy quarks. Right: Phase diagram for $N_f=2, N_\tau=6$ 
from the 3d effective theory \cite{fromm}.}
\label{fig:heavy}       % Give a unique label
\end{figure}

After successful comparison with the full 4d theory, one can switch on a finite chemical potential and study how
the finite temperature deconfinement transition changes. 
%The sign problem of the effective theory is mild enough for it to be simulable 
%by reweighted ordinary Monte Carlo as well as complex Langevin without runaway solutions \cite{fromm,bind}. It thus 
This gives
predictions for the phase diagram of QCD with heavy quarks that are unavailable from full lattice QCD simulations.

The resulting phase diagram is shown 
in \fig\ref{fig:heavy}. The first-order deconfinement transition is weakened
by a real chemical potential and disappears in a critical end point, which has been calculated as a function of pion mass \cite{fromm}.
This qualitative behaviour of the phase transition is reproduced by continuum studies using a Polyakov loop model \cite{lo} and in the  
functional renormalisation group approach \cite{frg}.

\section{The cold and dense regime}

The most difficult region to address is that of cold and dense QCD, since the sign problem grows exponentially with $\mu/T$.  
In order to understand the qualitative features of this region, it is instructive to consider 
the strong coupling ($\beta=0$) limit of the effective theory with a static quark determinant only. In this case the partition function 
factorises into one-site integrals which can be solved analytically. In the zero temperature limit,
mesonic contributions are exponentially suppressed by chemical potential and for  
$N_f=1$ we have  \cite{silver, bind} 
\beq
Z(\beta=0) \stackrel{T\rightarrow 0}{\longrightarrow}z_0^V \quad \mbox{with}\quad 
z_0=1+4h_1^{3}+h_1^{6}, \quad h_1=(2\kappa e^{a\mu})^{N_\tau}=e^{(m-\mu)/T}\;.
\eeq
Note that this corresponds to a free baryon gas with two species. With one quark flavour only, there are no nucleons and the
first prefactor indicates a spin 3/2  quadruplet of $\Delta$'s whereas the second term is a spin 0 six quark state or di-baryon.
The quark number density is now easily evaluated
\beq
n=
\frac{T}{V}\frac{\partial}{\partial \mu}\ln Z=\frac{1}{a^3}\frac{4N_ch_1^{N_c}+2N_ch_1^{2N_c}}{1+4h_1^{N_c}+h_1^{2N_c}}\;,
 \quad \lim_{T\rightarrow 0} a^3n=\left\{\begin{array}{cc} 0, & \mu<m\\
	2N_c, & \mu>m\end{array}\right.\;,
\eeq
and at zero temperature exhibits a discontinuity when the quark chemical potential equals the constituent mass $m$.
This reflects the ``silver blaze'' property of QCD, i.e.~the fact that the baryon number stays zero
for small $\mu$ even though the partition function explicitly depends on it \cite{cohen}. Once the baryon chemical potential 
$\mu_B=3 \mu$
is large enough to make a baryon ($m_B=3m$ in the static strong coupling limit), a discontinuous phase transition 
to a saturated baryon crystal takes place. 
Note that saturation density here is $2N_c$ quarks per flavour and lattice
site and reflects the Pauli principle. This is clearly a discretisation effect that has to disappear
in the continuum limit.

In the case of two flavours the corresponding expression for the free gas of baryons reads
\begin{eqnarray}
z_0& =& (1 + 4 h_d^3 + h_d^6)+ (6 h_d^2 + 4 h_d^5) h_u+ (6 h_d + 10 h_d^4)h_u^2+ 
  (4 + 20 h_d^3 + 4 h_d^6)h_u^3 \nn \\
&&  + (10 h_d^2 + 6 h_d^5) h_u^4+ ( 4 h_d + 6 h_d^4) h_u^5 
  +(1 + 4 h_d^3 + h_d^6)h_u^6\;,
\label{eq:freegas}  
\end{eqnarray}
where we have now distinguished between the $h_1$ coupling for the $u$- and $d$-quarks. In this case we easily identify in addition the 
spin 1/2 nucleons as well has many other baryonic multi-quark states with their correct spin degeneracy. A similar result is obtained for
mesons if we instead consider an isospin chemical potential in the low temperature limit \cite{bind}. Remarkably, the entire spin-flavour-structure
of the QCD bound states is obtained in this simple limit.

\begin{figure}[t]
\centering
\includegraphics[width=7cm]{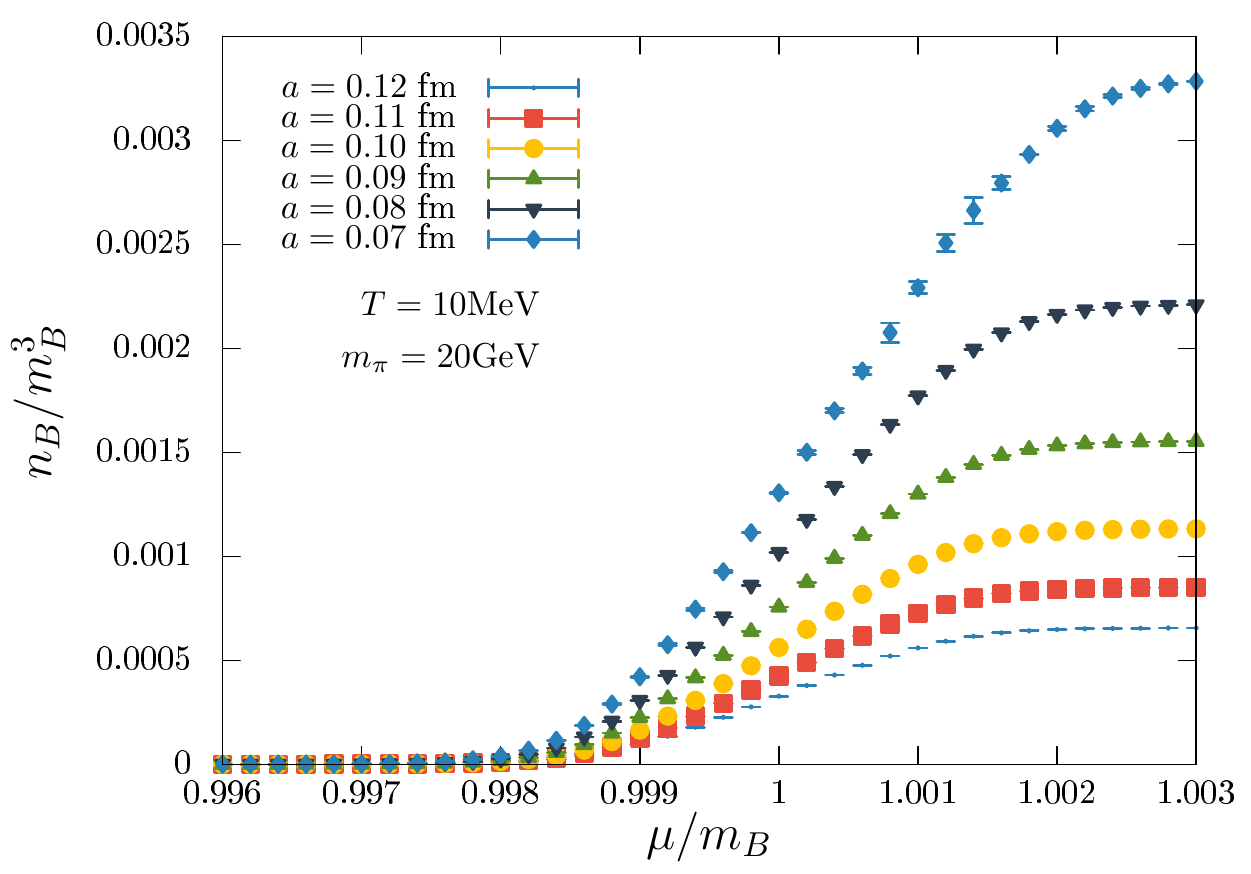}
\includegraphics[width=7cm,clip]{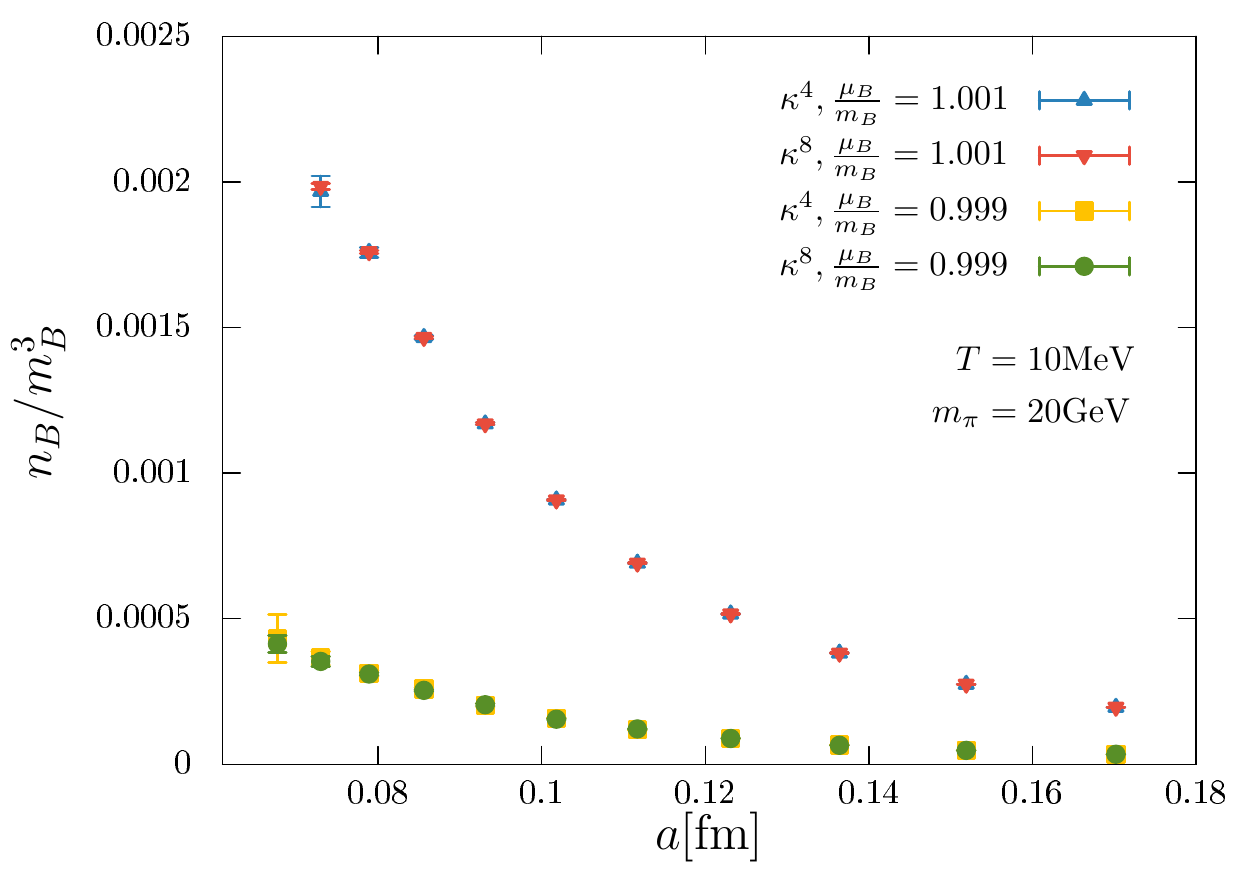}
\caption{Left: Baryon density in continuum units for various lattice spacings. Right: Continuum approach of the baryon density
for fixed chemical potentials. From \cite{k8}.}
\label{fig:cont}       % Give a unique label
\end{figure}

When corrections are taken into account, the step function transition gets smeared out. With suffiently many corrections at hand, also the
lattice spacing can be varied and the approach to the continuum can be studied, as in \fig\ref{fig:cont}, which is an analytic evaluation
through orders $u^5\kappa^8$ for a sufficiently heavy quark mass to reach the continuum with the calculated orders \cite{k8}. 
On the left we see how the 
lattice saturation of the baryon density, when plotted in physical units, grows with decreasing lattice spacing and disappears in the continuum.
At the same time, the plot illustrates the enormous difficulties for a continuum limit, even when there is no sign problem: because of the 
artefact of lattice saturation, a continuum extrapolation as in \fig\ref{fig:cont} (right) is rapidly harder to control with increasing density. 
Higher densities require finer lattices, $a\ll \mu^{-1}$.

Despite these difficulties, the calculations do reproduce features of continuum physics and give important insights. 
With interactions, the onset transition
is shifted to slightly smaller chemical potentials, as expected for a weak attractive interaction between baryons causing their condensation.
Further, the onset transition in the figure is a smooth crossover and not a first-order transition as expected for sufficiently low temperatures
and physical quark masses. The reason for this behaviour is indeed the large quark mass. At zero temperature the binding energy per baryon in 
units of the lowest baryon mass can be extracted from the following  thermodynamic quantity, which to leading order 
is proportional to $\kappa^2$ and therefore decreases with growing quark masses,
\beq
\epsilon= \frac{e-n_Bm_B}{n_Bm_B} =  -\frac{4}{3}\frac{1}{a^3n_B}\left(\frac{6h_1^3+3h_1^6}{z_0}\right)^2\,\kappa^2+\ldots\;.
\label{eq:bindpt}
\eeq
The binding energy per nucleon sets the scale for the temperature of the critical endpoint of the nuclear liquid gas transition, above which the nuclear
``liquid'' has completely evaporated. Indeed, calculations  for large values of $\kappa$ confirm that the liquid gas transition 
becomes first order at large $N_\tau$ (low T) and crossover at lower $N_\tau$ (higher T), implying the presence of a critical endpoint \cite{bind}.
Thus, our effective lattice theory derived from QCD shows all qualitative features of the physical nuclear liquid gas transition. 

\section{QCD for large $N_c$}

Since we have analytic control, both over the derivation as well as the evaluation of the effective theory, it is intriguing to investigate
what happens when the number of colours $N_c$ is varied and made large. In particular, we wish to explore possible contact to the 
large $N_c$ considerations leading to  the prediction of quarkyonic matter \cite{quarky}.

There is a lot of interesting literature on QCD at large $N_c$ which we are unable to represent properly. Here we just   
summarise the most essential features, most of them established in the early works \cite{hooft,witten}.
The limit of large $N_c$ of $SU(N_c)$-QCD is defined by 
\beq
N_c\rightarrow \infty\quad \mbox{with} \quad g^2N_c={\rm const.}
\label{eq:limit}
\eeq
In this case the theory has the following properties:
\begin{itemize}
\item Quark loops in Feynman diagrams are suppressed by $N_c^{-1}$
\item Non-planar Feynman diagrams are suppressed by $N_c^{-2}$
\item Mesons are free; the leading corrections are cubic interactions $\sim N_c^{-1/2}$ and quartic interactions $\sim N_c^{-1}$
\item Meson masses are $\sim \Lambda_{QCD}$
\item Baryons consist of $N_c$ quarks, baryon masses are $\sim N_c \Lambda_{QCD}$
\item Baryon interactions are $\sim N_c$
\end{itemize}

\begin{figure}[t]
\centering
\includegraphics[width=7cm]{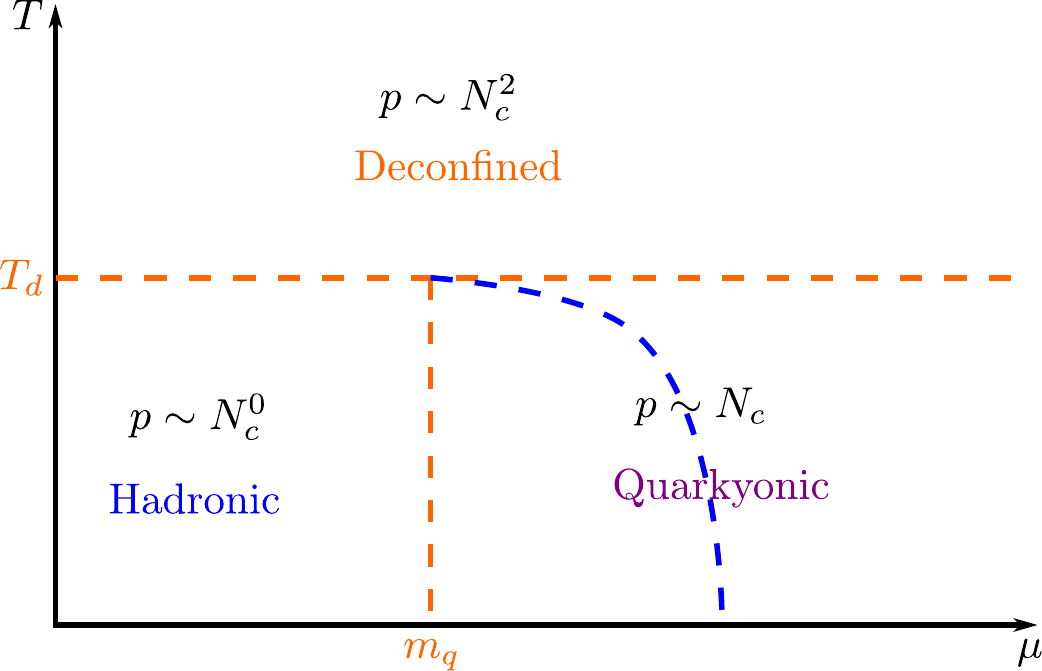}
\includegraphics[width=7cm]{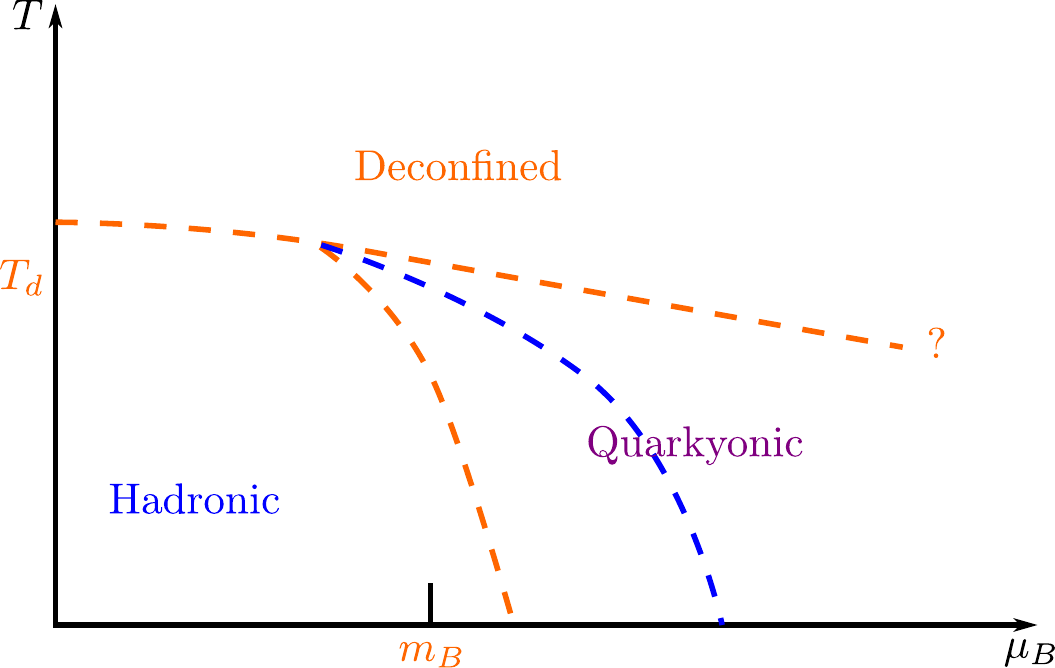}
\caption[]{Phase diagram in the limit of large $N_c$ (left) and possible consequences for $N_c=3$ (right) according to \cite{quarky}.
The blue line indicates the chiral transition. }
\label{fig:pd_nc}       % Give a unique label
\end{figure}

The authors of \cite{quarky} used these and various other ingredients to draw conclusions for the QCD phase diagram.
\fig\ref{fig:pd_nc} (left) shows the phase diagram in the large $N_c$ limit. With quark loops suppressed, the phase boundary 
of the deconfinement transition is unaffected by chemical potential, forming a horizontal line. Statistical mechanics
then says that in the hadronic, low density phase, the baryonic contribution to the pressure is exponentially suppressed 
with their mass, so $p\sim N_c^0$ there. In the plasma phase perturbation theory tells us that $p\sim N_c^2$. In \cite{quarky}
similar arguments for large $\mu$ suggest that at low temperatures and $\mu_B>m_B$, 
the pressure scales as $p\sim N_c$. The authors termed this phase ``quarkyonic'', since it shows both quark-like and baryon-like aspects,
for details we refer to the literature. 
Here we are interested in what we can say about this region based on the effective lattice theory.

\section{The effective theory for general $N_c$}

We now wish to generalise our effective theory to arbitrary $N_c$. Note that $N_c=2$ has already been 
analysed in detail \cite{scior}, with interesting physics results for two-colour QCD. 
Our aim here is to go in the other direction and to increase $N_c$.
Before we do so, we need to mention a few important caveats concerning large $N_c$, 
some of which we hope to address in the near future.
First, our current treatment can not yet follow the prescription (\ref{eq:limit}) of keeping the t'Hooft coupling constant,
since we work in the strong coupling limit and thus cannot vary $g^2$. 
This can be improved once we include gauge corrections. Second, there is the important question if and under which circumstances
the strong coupling and large $N_c$ limits commute \cite{gw}. 
The following discussion, in which we recalculate our previous results for general $N_c$, thus only represents a first step in this direction.

\subsection{Integration over $SU(N_c)$}

Rather than expanding for the leading behaviour at large $N_c$, we have been able to generalise 
the gauge integrations over the link variables to integration over general $SU(N_c)$ groups, by combining various results known in 
the literature. We consider class functions of $U(N_c)$ group elements, $f(U)=f(VUV^{-1})$, which are invariant under a change of basis.
These functions only depend on the eigenvalues $z_i$ of a group element and furthermore for our purposes it is sufficient to specialise to functions which factorise with respect to the eigenvalues,
\beq
f(U)= \tilde{f}(z_1, \ldots, z_{N_c}) = \tilde{f_1}(z_1) \cdot \ldots \cdot \tilde{f_{N_c}}(z_{N_c})\;.
\eeq 
For such functions the integration over the group can be expressed as \cite{Nishida:2003fb}
\begin{equation}
  \label{eq:facorized-haar-integral}
  \int\limits_{U(N_c)} d{U} f(U) =
  \frac{1}{(2\pi i)^{N_c}}
  \det_{1\leq i,j\leq N_c}\Big(\oint\limits_{\gamma} d{z_i}\; \tilde{f_i}(z_i) z_i^{j-i-1}\Big)\;,
\end{equation}%
where $\gamma$ is a closed curve around the origin in the complex plane.
In the next step, this integration has to be extended from $U(N_c)$ to $SU(N_c)$, which can be done according to the formula \cite{Ravagli:2007rw}
\begin{equation}
  \int\limits_{SU(N_c)} d{U} f(U) = \sum_{q=-\infty}^\infty\,\,\int\limits_{U(N_c)} d{U} \det(U)^q f(U)\;.
\end{equation}
For our evaluation of the effective theory, to the leading order in the hopping expansion, we need the integration over the static determinant,
for which we obtain 
\begin{equation}
 z_0= \int\limits_{SU(N_c)} dU \det(1 + h_1 U)^{2 N_f} =
  \sum_{p=0}^{2 N_f}
  \det_{1\leq i,j \leq N_c}\left(\binom{2 N_f}{i-j+p}\right)
  h_1^{p N} \;,
\end{equation}
where the matrix elements are given in terms of binomial coefficients specified by $i,j$. The corresponding determinant evaluates to
\begin{align}
  \det_{1\leq i,j \leq N}\left(\binom{2 N_f}{i-j+p}\right)
%  & = \prod_{i=1}^N \frac{(i-1+2 N_f)^{\underline{p}}}{(i-1+p)^{\underline{p}}} \\
  & = \prod_{i=1}^p \frac{(i-1+2 N_f-p+N)^{\underline{2 N_f-p}}}{(i-1+2 N_f-p)^{\underline{2 N_f-p}}}\;,
\end{align}
where the underline notation indicates falling factorials, $n^{\underline{k}} = n \cdot (n-1) \cdots (n-k+1)$.
To evaluate the first correction in the hopping expansion, we need a further integal for which we find
\begin{equation}
  \int\limits_{SU(N_c)} dU \det(1+h_1 U)^{2 N_f} \tr\left(\frac{h_1 U}{1 + h_1 U}\right) =
  \sum_{p=0}^{2 N_f} \det_{1\leq i,j\leq N_c}\left(\binom{2 N_f}{i-j+p}\right) \frac{p N_c}{2 N_f} h_1^{p N_c}\;.
\end{equation}

\subsection{Results for general $N_c$}

Evaluating the integrals according to the last section, we find for the static determinant and the first correction
in the strong coupling limit of the $SU(N_c)$-theory with $N_f=2$ flavours:
\begin{align}
  z_0
  & = 1 + \frac{1}{6}(h_1^{N_c}+h_1^{3 N_c})(N_c+3)(N_c+2)(N_c+1) +
    \frac{1}{12}h_1^{2 N_c}(N_c+3)(N_c+2)^2(N_c+1) + h_1^{4 N_c}\;, \\
  z_{11}
  & = \frac{1}{24}h_1^{N_c}(N_c+3)(N_c+2)(N_c+1)N_c + \frac{1}{24}h_1^{2 N_c}(N_c+3)(N_c+2)^2(N_c+1)N_c\nn \\
  & \phantom{{}=} + \frac{1}{8}h_1^{3 N_c}(N_c+3)(N_c+2)(N_c+1)N_c + h_1^{4 N_c} N_c\;.
\end{align}
In particular, the first expression is to be compared with the special case for $N_c=3$ in (\ref{eq:freegas}) and reveals an
interesting insight.
The spin degeneracy prefactors of the free baryon gas are determined by $N_c$, which is of course
due to the fact that the spins of the $N_c$ quarks in a baryon combine to a baryon spin.  

\begin{figure}[t]
\centering
\includegraphics[width=7.5cm]{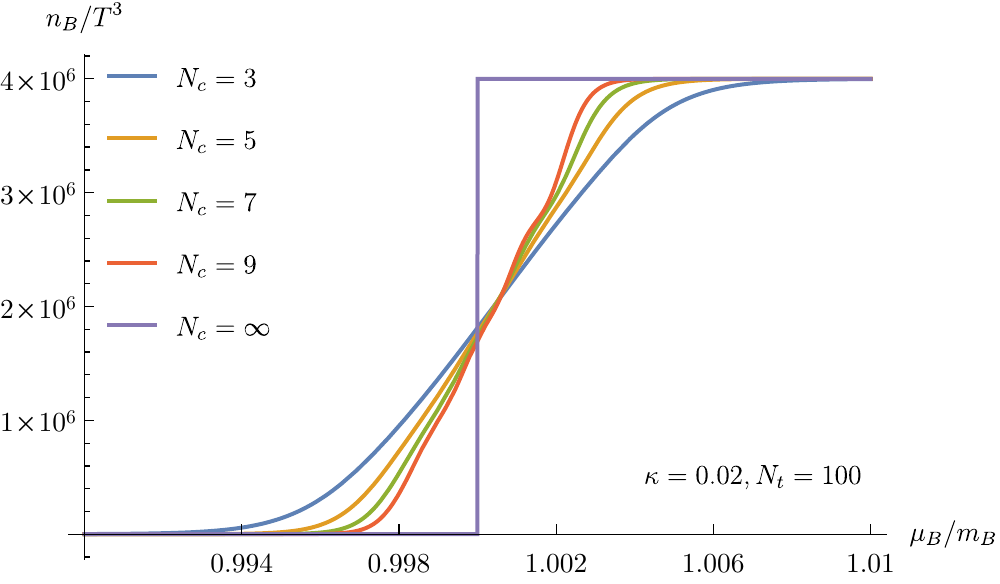}
\includegraphics[width=7cm]{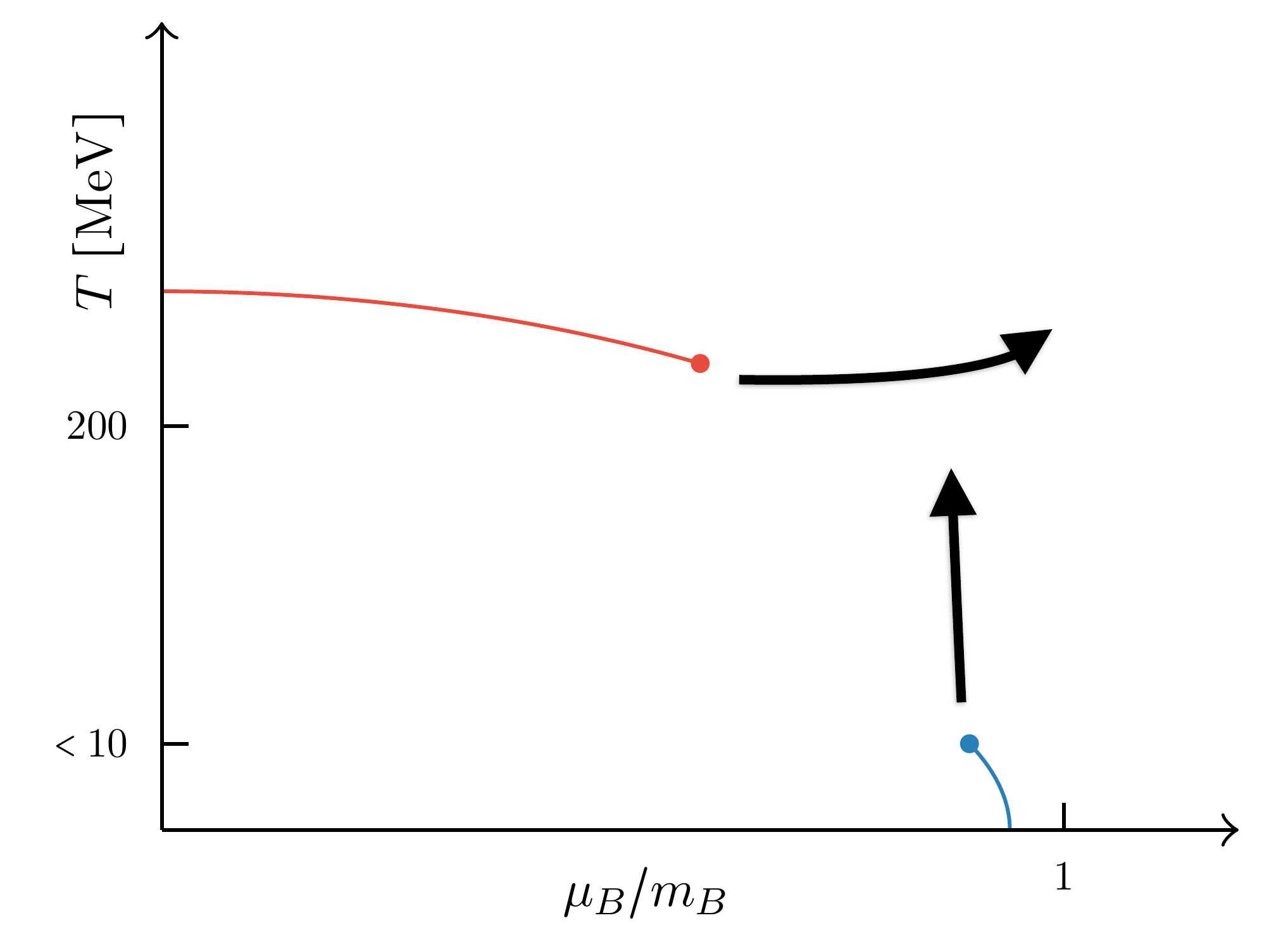}
\caption[]{Left: Onset transition for different values of $N_c$. Right: Arrows indicate the smooth change of the phase transition lines 
with growing $N_c$.}
\label{fig:onset_nc}       % Give a unique label
\end{figure}

We can now evaluate the thermodynamic functions entering the equation of state for any desired value of $N_c$. 
The free energy density is in our approximation given by the static limit plus a correction term, 
\begin{equation}
  -f = \log(z_0) + \frac{\kappa^2 N_{\tau}}{N_c}(-6 N_f)\left(\frac{z_{11}}{z_0}\right)^2,
\end{equation}
and from it all others can be easily calculated.
\fig\ref{fig:onset_nc} (left) shows once more the onset transition to finite baryon density for different choices of $N_c$, 
up until lattice saturation is reached. 
We observe a steepening of the transition with increasing $N_c$, which asymptotically ends up in a step function, i.e.~ a 
first-order transition, even though we started with a smooth crossover at $N_c=3$. Thus, growing 
$N_c$ appears to make the onset transition to baryon matter more strongly first-order. This implies that the temperature of the transition's critical
endpoint has to increase with growing $N_c$ until it hits another discontinuity, as indicated in \fig\ref{fig:onset_nc} (right). 
Since we know already that also the deconfinement transition line has to ``straighten out'' with growing $N_c$, we observe how the rectangular
phase diagram of \fig\ref{fig:pd_nc} (left) emerges smoothly from \fig\ref{fig:onset_nc} (right) by increasing $N_c$.

\subsection{Large $N_c$}

Next we extract the terms dominating the thermodynamic functions at large $N_c$. In this analysis we have to carefully 
distinguish between the leading and subleading term in the hopping expansion. Beyond the onset of baryon condensation, 
the leading term represents the lattice saturation, which is an unphysical artefact of discretisation. Interestingly, correction terms 
do {\it not} contribute to saturation, but rather modify the smooth shape of the curves entering their low and high density asymptotes.
We then analyse the behaviour of these terms to the left and the right of the onset transition.
\bea
\mu<m, h_1<1:&\quad  a^4 p_0 \sim \frac{1}{6 N_{\tau}} N_c^3 h_1^{N_c},\quad &\quad a^4 p_1 \sim -\frac{1}{48} N_c^7 h_1^{2 N_c} \\
&\quad a^3 n_0 \sim \frac{1}{6} N_c^4 h_1^{N_c} \quad & \quad a^3 n_1 \sim \frac{N_{\tau}}{24} N_c^9 h_1^{2 N_c}\\
&& \nonumber \\
\mu>m,h_1>1:&\quad a^4 p_0 \sim \frac{4 \log(h_1)}{N_{\tau}} N_c\quad & \quad a^4 p_1 \sim -12 N_c\\
&  \quad  a^3 n_0 \sim 4 N_c \quad & \quad a^3 n_1  \sim -288 N_{\tau} \frac{N_c^5}{h_1^{N_c}}
\eea
For chemical potentials below the onset transition, the fugacity $h_1$ to the power of some $N_c$ always overwhelms the prefactor, 
leading to an exponential suppression of both pressure and baryon density, thus pushing the silver blaze phenomenon further to the right,
causing part of the observed steepening of the onset transition. Note, how our calculation fully reproduces the physics argument based
on the statistical mechanics of a weakly interacting baryon gas. For chemical potentials larger than the onset transition, the analysis is more tricky.
The first term, representing lattice saturation due to Pauli blocking, trivially scales with $N_c$, but is unphysical and should disappear to
infinity in the continuum, as discussed above. Intriguingly however, we also find the first correction term to the pressure to have a leading
linear $N_c$ dependence. This term does {\it not} end up in lattice saturation and thus contributes to continuum physics. 
Taken at face value, this would suggest that, immediately after the onset transition, quarkyonic matter and baryonic matter are the same.
Clearly, this preliminary investigation should be supplemented by gauge and possibly higher order corrections, as well as an analysis of the possible  caveats mentioned before. 

\section{Conclusions}

We have derived and evaluated an effective 3d lattice theory, which is capable to address the thermodynamics of 
the cold and dense region of QCD with heavy quarks, in particular the onset transition to baryon matter. 
The effective theory predicts a first order deconfinement transition terminating in a critical end point, and a nuclear liquid
gas transition terminating in a critical end point. 
We have shown preliminary results of employing this effective theory to QCD with a general number of colours, $N_c$.
In the strong coupling limit we find that the endpoint of the liquid gas transition moves to larger temperatures as $N_c$ is increased,
and becomes always first order when $N_c\rightarrow \infty$. 
Moreover, our analytic results indicate that the pressure to 
the right of the baryon onset scales as $p\sim N_c$, which is consistent with the definition of quarkyonic matter given in \cite{quarky}.  
This suggests the intriguing question whether baryonic and quarkyonic matter might be the same in this region of the phase diagram.
A necessary next step is the inclusion of gauge corrections and an analysis, how this affects the large $N_c$-behaviour 
when keeping $g^2N_c$ fixed.
\\
\\
\noindent
{\bf Acknowledgements}
This work was partially funded by the Deutsche Forschungsgemeinschaft 
(DFG, German Research Foundation), project number 315477589 -TRR 211. We also acknowledge support 
by the Helmholtz International Center for FAIR within the LOEWE program of the State of Hesse.


\begin{thebibliography}{99}

\bibitem{review}
P.~de Forcrand,
  %``Simulating QCD at finite density,''
  PoS LAT {\bf 2009}, 010 (2009),
  [arXiv:1005.0539 [hep-lat]].
  %%CITATION = ARXIV:1005.0539;%%
\bibitem{cf}
C.~S.~Fischer,
  %``QCD at finite temperature and chemical potential from Dyson-Schwinger equations,''
  arXiv:1810.12938 [hep-ph].
  %%CITATION = ARXIV:1810.12938;%%
  \bibitem{quarky}
L.~McLerran and R.~D.~Pisarski,
  %``Phases of cold, dense quarks at large N(c),''
  Nucl.\ Phys.\ A {\bf 796} (2007) 83
  doi:10.1016/j.nuclphysa.2007.08.013
  [arXiv:0706.2191 [hep-ph]].
  %%CITATION = doi:10.1016/j.nuclphysa.2007.08.013;%% 
 \bibitem{mish}
 G.~Torrieri, S.~Lottini, I.~Mishustin and P.~Nicolini,
  %``The Phase diagram in $T-\mu-N_c$ space,''
  Acta Phys.\ Polon.\ Supp.\  {\bf 5} (2012) 897
  doi:10.5506/APhysPolBSupp.5.897
  [arXiv:1110.6219 [nucl-th]].
  %%CITATION = doi:10.5506/APhysPolBSupp.5.897;%% 
 \bibitem{wozar}
 C.~Wozar, T.~Kaestner, A.~Wipf and T.~Heinzl,
  %``Inverse Monte-Carlo determination of effective lattice models for SU(3) Yang-Mills theory at 
  %  finite temperature,''
  Phys.\ Rev.\ D {\bf 76}, 085004 (2007) 
  [arXiv:0704.2570 [hep-lat]].
  %%CITATION = ARXIV:0704.2570;%%    
  \bibitem{green1}
J.~Greensite and K.~Langfeld,
  %``Effective Polyakov line action from strong lattice couplings to the deconfinement transition,''
  Phys.\ Rev.\ D {\bf 88}, 074503 (2013) 
  [arXiv:1305.0048 [hep-lat]].
%\cite{Greensite:2014isa}
\bibitem{green2}
  J.~Greensite and K.~Langfeld,
  %``Finding the effective Polyakov line action for SU(3) gauge theories at finite chemical potential,''
  Phys.\ Rev.\ D {\bf 90},  014507 (2014)
  [arXiv:1403.5844 [hep-lat]].
  %%CITATION = ARXIV:1403.5844;%%
\bibitem{bergner}
G.~Bergner, J.~Langelage and O.~Philipsen,
  %``Numerical corrections to the strong coupling effective Polyakov-line action for finite T Yang-Mills theory,''
  JHEP {\bf 1511}, 010 (2015)
 % doi:10.1007/JHEP11(2015)010
  [arXiv:1505.01021 [hep-lat]].
  %%CITATION = doi:10.1007/JHEP11(2015)010;%%  
\bibitem{polon}
J.~Polonyi and K.~Szlachanyi,
  %``Phase Transition from Strong Coupling Expansion,''
  Phys.\ Lett.\  {\bf 110B} (1982) 395.
  doi:10.1016/0370-2693(82)91280-1
  %%CITATION = doi:10.1016/0370-2693(82)91280-1;%%  
  \bibitem{efft1}
J.~Langelage, S.~Lottini and O.~Philipsen,
  %``Centre symmetric 3d effective actions for thermal SU(N) Yang-Mills from strong coupling series,''
  JHEP {\bf 1102}, 057 (2011) 
   [JHEP {\bf 1107}, 014 (2011)]
  [arXiv:1010.0951 [hep-lat]].
  %%CITATION = ARXIV:1010.0951;%%   
\bibitem{lange1}
  J.~Langelage, G.~M\"unster and O.~Philipsen,
  %``Strong coupling expansion for finite temperature Yang-Mills theory in the confined phase,''
  JHEP {\bf 0807}, 036 (2008)
%  doi:10.1088/1126-6708/2008/07/036
  [arXiv:0805.1163 [hep-lat]].
  %%CITATION = doi:10.1088/1126-6708/2008/07/036;%%
  \bibitem{lange2}
J.~Langelage and O.~Philipsen,
  %``The deconfinement transition of finite density QCD with heavy quarks from strong coupling series,''
  JHEP {\bf 1001},  089 (2010)
  [arXiv:0911.2577 [hep-lat]].
  %%CITATION = ARXIV:0911.2577;%%
  \bibitem{sy}
B.~Svetitsky and L.~G.~Yaffe,
  %``Critical Behavior at Finite Temperature Confinement Transitions,''
  Nucl.\ Phys.\ B {\bf 210} (1982) 423.
  doi:10.1016/0550-3213(82)90172-9
  %%CITATION = doi:10.1016/0550-3213(82)90172-9;%%
\bibitem{fromm}
M.~Fromm, J.~Langelage, S.~Lottini and O.~Philipsen,
  %``The QCD deconfinement transition for heavy quarks and all baryon chemical potentials,''
  JHEP {\bf 1201} (2012) 042
  doi:10.1007/JHEP01(2012)042
  [arXiv:1111.4953 [hep-lat]].
  %%CITATION = doi:10.1007/JHEP01(2012)042;%%
\bibitem{bind}
J.~Langelage, M.~Neuman and O.~Philipsen,
  %``Heavy dense QCD and nuclear matter from an effective lattice theory,''
  JHEP {\bf 1409}, 131 (2014) 
  [arXiv:1403.4162 [hep-lat]].
  %%CITATION = ARXIV:1403.4162;%%      
 \bibitem{k8}
J.~Glesaaen, M.~Neuman and O.~Philipsen,
  %``Equation of state for cold and dense heavy QCD,''
  JHEP {\bf 1603}, 100 (2016)
%  doi:10.1007/JHEP03(2016)100
  [arXiv:1512.05195 [hep-lat]].
  %%CITATION = doi:10.1007/JHEP03(2016)100;%%  
 \bibitem{lo}
P.~M.~Lo, B.~Friman and K.~Redlich,
  %``Polyakov loop fluctuations and deconfinement in the limit of heavy quarks,''
  Phys.\ Rev.\ D {\bf 90}, no. 7, 074035 (2014)
%  doi:10.1103/PhysRevD.90.074035
  [arXiv:1406.4050 [hep-ph]].
  %%CITATION = doi:10.1103/PhysRevD.90.074035;%%
 \bibitem{frg}
 C.~S.~Fischer, J.~Luecker and J.~M.~Pawlowski,
  %``Phase structure of QCD for heavy quarks,''
  Phys.\ Rev.\ D {\bf 91}, no. 1, 014024 (2015)
%  doi:10.1103/PhysRevD.91.014024
  [arXiv:1409.8462 [hep-ph]].
  %%CITATION = doi:10.1103/PhysRevD.91.014024;%% 
\bibitem{silver}
M.~Fromm, J.~Langelage, S.~Lottini, M.~Neuman and O.~Philipsen,
  %``Onset Transition to Cold Nuclear Matter from Lattice QCD with Heavy Quarks,''
  Phys.\ Rev.\ Lett.\  {\bf 110} no. 12, 122001 (2013)  
  [arXiv:1207.3005 [hep-lat]].
  %%CITATION = ARXIV:1207.3005;%%
 \bibitem{cohen}
T.~D.~Cohen,
  %``Functional integrals for QCD at nonzero chemical potential and zero density,''
  Phys.\ Rev.\ Lett.\  {\bf 91}, 222001 (2003)
  doi:10.1103/PhysRevLett.91.222001
  [hep-ph/0307089].
  %%CITATION = doi:10.1103/PhysRevLett.91.222001;%%  
\bibitem{hooft}
G.~'t Hooft,
  %``A Planar Diagram Theory for Strong Interactions,''
  Nucl.\ Phys.\ B {\bf 72} (1974) 461.
  doi:10.1016/0550-3213(74)90154-0
  %%CITATION = doi:10.1016/0550-3213(74)90154-0;%%
\bibitem{witten}
E.~Witten,
  %``Baryons in the 1/n Expansion,''
  Nucl.\ Phys.\ B {\bf 160} (1979) 57.
  doi:10.1016/0550-3213(79)90232-3
  %%CITATION = doi:10.1016/0550-3213(79)90232-3;%%    
\bibitem{scior}
P.~Scior and L.~von Smekal,
  %``Baryonic Matter Onset in Two-Color QCD with Heavy Quarks,''
  Phys.\ Rev.\ D {\bf 92} (2015) no.9,  094504
  doi:10.1103/PhysRevD.92.094504
  [arXiv:1508.00431 [hep-lat]].
  %%CITATION = doi:10.1103/PhysRevD.92.094504;%%  
\bibitem{gw}
D.~J.~Gross and E.~Witten,
  %``Possible Third Order Phase Transition in the Large N Lattice Gauge Theory,''
  Phys.\ Rev.\ D {\bf 21} (1980) 446.
  doi:10.1103/PhysRevD.21.446
  %%CITATION = doi:10.1103/PhysRevD.21.446;%%  
\bibitem{Nishida:2003fb}
Y.~Nishida,
  %``Phase structures of strong coupling lattice QCD with finite baryon and isospin density,''
  Phys.\ Rev.\ D {\bf 69} (2004) 094501
  doi:10.1103/PhysRevD.69.094501
  [hep-ph/0312371].
  %%CITATION = doi:10.1103/PhysRevD.69.094501;%%
\bibitem{Ravagli:2007rw}
L.~Ravagli and J.~J.~M.~Verbaarschot,
  %``QCD in One Dimension at Nonzero Chemical Potential,''
  Phys.\ Rev.\ D {\bf 76} (2007) 054506
  doi:10.1103/PhysRevD.76.054506
  [arXiv:0704.1111 [hep-th]].
  %%CITATION = doi:10.1103/PhysRevD.76.054506;%%
  %28 citations counted in INSPIRE as of 30 Nov 2018
\end{thebibliography}
\end{document}